\documentclass[aip,rsi,reprint,pdftex,graphicx]{revtex4-1} 
\usepackage{amsmath,amssymb}
\usepackage[T1]{fontenc}
\usepackage[utf8]{inputenc}
\usepackage{graphicx}%
\usepackage{microtype}
\usepackage{textcomp}
\usepackage{xspace}
\usepackage{hyperref}

\newlength{\figwidth}
\newcommand{\cfeldesy}{\affiliation{Center for Free-Electron Laser Science, Deutsches
      Elektronen-Synchrotron DESY, Notkestraße 85, 22607 Hamburg, Germany}}%
\newcommand{\uhhphys}{\affiliation{Department of Physics, Universität Hamburg, Luruper Chaussee 149,
      22761 Hamburg, Germany}}%
\newcommand{\uhhcui}{\affiliation{Center for Ultrafast Imaging, Universität of Hamburg, Luruper
      Chaussee 149, 22761 Hamburg, Germany}}%
\newcommand{\stemail}{\email[]{sebastian.trippel@cfel.de}}%
\newcommand{\cmiweb}{\homepage{https://www.controlled-molecule-imaging.org}}%

\newcommand{\HHO}{\ensuremath{\text{H}_2\text{O}}\xspace}

\newcommand{\indolew}{\ensuremath{\text{indole}(\HHO)}\xspace}
\newcommand{\indolewd}{\ensuremath{\text{indole}(\HHO)_2}\xspace}
\newcommand{\indolewt}{\ensuremath{\text{indole}(\HHO)_3}\xspace}
\newcommand{\indoleD}{\ensuremath{(\text{indole})_2}\xspace}
\newcommand{\indoleDw}{\ensuremath{(\text{indole})_2(\HHO)}\xspace}
\newcommand*{\ordsim}{\mathord{\sim}}
\newcommand*{\eg}{e.\,g.}
\newcommand*{\celsius}[1]{\ensuremath{\xspace#1\,^\circ{}\text{C}}\xspace}
\newcommand*{\um}{\ensuremath{\text{\textmu{m}}}\xspace}%

\begin{document}
\title{NOTE: Knife edge skimming for improved separation of molecular species by the deflector}
\author{Sebastian Trippel}\stemail\cmiweb\cfeldesy\uhhcui
\author{Melby Johny}\thanks{These authors contributed equally.}\cfeldesy\uhhcui
\author{Thomas Kierspel}\thanks{These authors contributed equally.}\cfeldesy\uhhcui\uhhphys
\author{Jolijn Onvlee}\cfeldesy
\author{Helen Bieker}\cfeldesy\uhhcui
\author{Hong Ye}\cfeldesy\uhhcui\uhhphys
\author{Terry Mullins}\cfeldesy
\author{Lars Gumprecht}\cfeldesy
\author{Karol Długołęcki}\cfeldesy
\author{Jochen Küpper}\cfeldesy\uhhcui\uhhphys
\date{\today}
\begin{abstract}
   A knife edge for shaping a molecular beam is described to improve
   the spatial separation of the species in a molecular beam by the
   electrostatic deflector. The spatial separation of different
   molecular species from each other as well as from atomic seed gas
   is improved. The column density of the selected molecular-beam part
   in the interaction zone, which corresponds to higher signal rates,
   was enhanced by a factor of 1.5, limited by the virtual source size
   of the molecular beam.
\end{abstract}
\maketitle

Molecular-beam methods are important in physical chemistry and
molecular physics, as they provide unique opportunities to obtain
fundamental insight into mechanisms and dynamics of elementary
molecular and chemical processes. Furthermore, industrial applications
using molecular beams range from the fabrication of thin films to the
production of artificial structures such as quantum wires and dots.

Supersonic expansion of a gas into vacuum provides extreme cooling, in
the case of atomic or seeded molecular beams typically from ambient or
elevated temperatures down to
$\ordsim1$~K~\cite{Scoles:MolBeam:1and2}. This approach is used for a
large variety of experiments. In many applications the molecular beam
is shaped by skimmers, knife edges, razor blades, slits,
slit-skimmers, or gratings to select only the most intense part of the
beam~\cite{Gentry:RSI46:104, Scoles:MolBeam:1and2,
   Demtroeder:LaserSpectroscopy:1, Subramanian:RSI79:016101,
   Hornberger:RMP84:157}. Furthermore, molecular beams can be
manipulated by electric and magnetic fields which allow, \eg, the
separation of the molecules from a seed
gas~\cite{Meerakker:CR112:4828, Chang:IRPC34:557}.

Spatial separation of different species is achieved by the
electrostatic deflector~\cite{Ramsey:MolBeam:1956,
  Filsinger:JCP131:064309, Chang:IRPC34:557}. Experiments are in this
case typically performed at the edge of the deflected molecular beam
to maximize the separation or to reduce the amount of signal
originating from the seed gas. However, at this position of the beam
profile the column density of molecules is rather low. In this note,
we present the combination of a knife edge with the electrostatic
deflector, which allows for a better separation of the different
species of a molecular beam as well as an increase in column density
in the interaction region.

\begin{figure}
   \centering%
   \includegraphics[width=\linewidth]{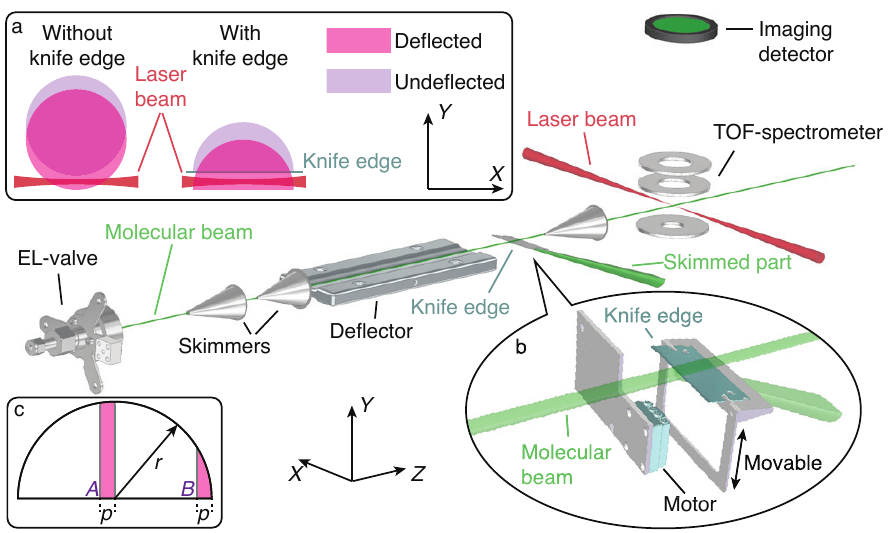}
   \caption{(Color online) Schematic of the experimental setup and the definition of the coordinate
      system. a) Sketch of the cross sections of the molecular beam and the laser beam
      to illustrate the working principle of the knife edge. b) Zoom into the knife edge region,
      showing the mechanical setup and motorization. c) Definition of the volumes $A$ and $B$, the
      beam radius $r$, and the width $p$ used for the theoretical limit; see text for details.}
   \label{fig:setup}
\end{figure}
A schematic of the experimental setup is shown in FIG.~\ref{fig:setup}. A pulsed molecular beam was
provided by expanding a few millibar of indole and a trace of water in 80~bar of helium through a
position-adjustable Even-Lavie valve~\cite{Even:JCP112:8068}. The valve was operated at a
temperature of \celsius{110} and at a repetition rate of 250~Hz. Two transversely, in $X$--$Y$,
adjustable conical skimmers (Beam Dynamics, model 50.8 with $\varnothing=3.0$, model 40.5 with
$\varnothing=1.5$~mm) were placed 6.5~cm and 30.2~cm downstream from the nozzle, respectively. The
transversely adjustable electrostatic deflector was located 4.4~cm behind the tip of the second
skimmer. Using the b-type electrostatic deflector~\cite{Kienitz:JCP147:024304}, the molecular beam
was dispersed according to the specific quantum states of the molecular
species~\cite{Chang:IRPC34:557, Filsinger:JCP131:064309, Trippel:PRA86:033202}. The vertically, $Y$,
adjustable knife edge was placed 1.3~cm behind the end of the deflector.

For the measurements with knife edge its vertical position was chosen such that the undeflected
molecular beam was cut roughly in its center. For the measurements without knife edge it was moved
vertically out of the molecular beam. A third, transversely adjustable skimmer (Beam Dynamics, model
50.8 with $\varnothing=1.5$~mm) was placed 2.5~cm downstream of the front of the knife edge. The
molecular beam entered a time of flight mass spectrometer (TOF-MS) centered 17.6~cm downstream of
the last skimmer, where the molecules and clusters were strong-field ionized by a laser pulse with a
pulse duration of 30~fs, centered at a wavelength of 800~nm, and focused to
$\varnothing\approx50$~\um. FIG.~\ref{fig:setup}{a} shows a cross section, in the $X$--$Y$ plane,
of the molecular beam to schematically illustrate the working principle of the knife edge. On the
left, a molecular beam profile defined by the shape of a round skimmer is depicted. Its deflected
part is shown by a vertical shift. On the right, the corresponding profiles are depicted for the
case with the knife edge. The laser probes the molecules in the deflected part of the beam,
resulting in a higher column density compared to the case without knife edge.
FIG.~\ref{fig:setup}{b} highlights the region of the setup where the knife edge was located. It
depicts the knife edge with its holder which was mounted on a motor (SmarAct SLC-1750-S-UHV) which
allows to position the knife edge vertically. The molecular beam is indicated by the green cylinder
which is cut into halves by the knife edge.

\begin{figure}
   \includegraphics[width=\linewidth]{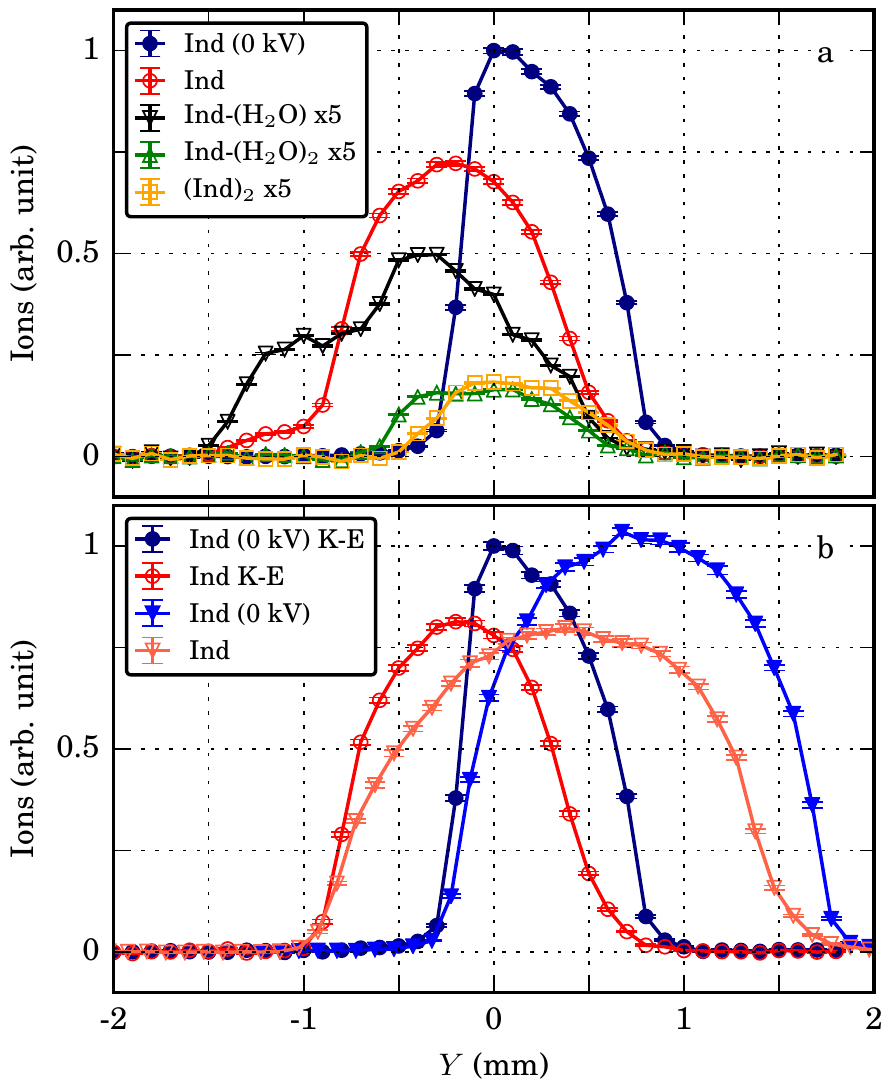}
   \caption{(Color online) a) Column density profile with knife edge of indole (dark blue), and
     deflection curves of indole (red open circles), \indolew (black open triangles), \indolewd
     (green open triangles), and \indoleD (orange open squares).  b) Column density profiles with
     deflector switched off without knife edge (blue triangles), deflector switched off with knife
     edge (dark blue circles), deflector switched on without knife edge (light red open triangles),
     and deflector switched on with knife edge (red open circles).}
   \label{fig:deflection-profile}
\end{figure}

We used the separation of indole and indole-water clusters to demonstrate the advantage of using the
knife edge in combination with the electrostatic deflector. FIG.~\ref{fig:deflection-profile}{a}
shows the measured vertical density profiles of the undeflected and deflected molecular beam when
the knife edge was used. The TOF mass spectrum was gated on specific masses, which corresponded to
either parent ions or specific fragments, to obtain each individual profile. The undeflected (0~kV)
profile of the signal corresponding to the indole mass of $m=117$~u is shown in dark blue. All
molecules and clusters were deflected downwards when voltages of $\pm10$~kV were applied to the
deflector electrodes, as all quantum states were high-field seeking at the electric field strengths
experienced inside the deflector~\cite{Chang:IRPC34:557}. The deflection profiles for the gates set
to the masses of indole, \indolew, \indolewd and \indoleD are shown in red, black, green, and
orange, respectively. The profiles for \indolew, \indolewd, and \indoleD were multiplied by a factor
of five. The \indolewt cluster was not observed in the mass spectrum. Furthermore, the profile of
\indoleDw had the same shape as the one for \indoleD and is not shown in the figure. Several edges
were observed in the profiles which correspond to various molecules and fragments. Going from left
to right, the outermost edge at -1.25~mm is attributed to \indolew because this cluster showed the
largest Stark effect of all molecules and clusters to be considered and was, therefore, deflected
the most~\cite{Trippel:PRA86:033202, Chang:IRPC34:557}. The shape of this edge matches the
corresponding edge in the indole-ion profile, which confirms that the \indolew ion was fragmenting
to indole ion with a probability of $\ordsim53$~\%. The edge at -0.9~mm in the indole-cation signal
was attributed to the indole monomer, since indole had the second largest Stark effect. The edge on
top of the \indolew profile at -0.6~mm was produced by \indolewd clusters which fragmented into
\indolew with a probability of $\ordsim64$~\%. A better separation of \indolew from indole and higher
clusters was observed in comparison to our previous experiments on this system without the knife
edge~\cite{Trippel:PRA86:033202, Chang:IRPC34:557}. Furthermore, the edge for the \indolewd cluster
has now been observed for the first time.

FIG.~\ref{fig:deflection-profile}{b} shows the measured deflection profiles for indole corrected
by the known fragmentation probabilities to account for the fragmentation for the case with and
without knife edge (Knife) and the deflector switched on (red) and off (blue). The profiles for the
case without knife edge were shifted by 0.975~mm to the left to match the edges on the left side for
a better direct comparison. In both cases -- deflector on and off -- the left edge was steeper for
the measurements with knife edge. This is attributed to the higher column density as a result of the
knife edge. Placing the probe laser at $-0.7$~mm in the deflected profile results in an enhancement
factor of $R=1.5$ at this position. The measured molecular beam diameter of $r=2$~mm matches exactly
the expected radius from geometry arguments assuming a point source for the molecular beam. The
distances between the valve and the third skimmer and the interaction region are 53.4~cm and 71~cm,
respectively. This results in a magnification factor of $71.0/53.4=1.33$, in excellent agreement
with the ratio between the measured molecular-beam diameter and the skimmer diameter given by
$2.0/1.5=1.33$. The deflected part of the molecular beam is, therefore, also expected to be far out
of the geometric helium profile. Additional broadening mechanisms for the molecular beam, such as
the finite temperature or deviations from a point source are not taken into account. The influence
of these contributions to the purity of the molecular beam are beyond the scope of this manuscript.

The maximum enhancement factor $R$ for the increase in column density can be estimated, assuming an
uniform molecular beam emitted from a point source and a uniform deflection force, from the
molecular beam radius $r$ in the interaction region and the width of the, for the interaction with
the molecules relevant, volume $p$. For $p\ll{}r$ the enhancement factor is given by
$R=A/B=3/4\sqrt{2r/p}$, see FIG.~\ref{fig:setup}{c}. Taking the radius of our measured molecular
beam profile of $r=1.0$~mm and the diameter of the ionization laser $p=50$~\um resulted in an
expected enhancement factor of $R\approx4.7$. We attribute the reduced experimentally observed
enhancement factor of $R=1.5$ to the following reasons: The experimental molecular beam profile was
not completely collimated and, therefore, the edges of the profiles are not infinitely steep. This
is ascribed to the finite size of the virtual source, which we estimate to be in the order of
0.6~mm. Due to the finite source size and the geometrical constraints given by the skimmers we
furthermore expect it be be advantageous to place the knife edge behind the deflector, compared to
using, \eg, a slit-skimmer before the deflector since it decreases the effective virtual source.
Secondly, the important volume for the interaction of the molecules with the ionization laser was
unknown and might be broader than the measured diameter in intensity. A third contribution to the
reduced enhancement is attributed to the fact that the deflector acts as a thick lens for the
dispersion of the molecular beam which leads to a softening of the edges. A further contribution
could be a misalignment of the knife edge with respect to the propagation direction of the probe
laser.

The combination of the knife edge with the electrostatic deflector is of general use for all
molecular beam experiments that benefit from a strong separation of molecular species or a strong
separation from the seed gas. The presented approach is also especially useful for applications with
low count rates or restricted measurement times, \eg, beamtimes at large facilities such as free
electron lasers (FELs), synchrotrons, or high-power-laser facilities, where typically only a few
days of beamtime are available for the measurements. Furthermore, for probing a collimated molecular
beam with $r=2$~mm by the generally small x-ray beams, \eg, $p=5$~\um, a theoretical enhancement
factor, according to the model described above, of $R>20$ is obtained. Taking into account the
finite virtual source size and the resulting measured reduction of the enhancement factor by about
$5/1.5$ leads to an expected enhancement factor of 6 in line with preliminary results from a recent
beamtime at the LCLS.

\begin{acknowledgments}
We acknowledge Benjamin Erk and the CAMP team for a significant equipment loan.

Besides DESY, this work has been supported by the excellence cluster ``The Hamburg Center for
Ultrafast Imaging -- Structure, Dynamics and Control of Matter at the Atomic Scale'' (CUI,
DFG-EXC1074), the European Research Council under the European Union's Seventh Framework Programme
(FP7/2007-2013) through the Consolidator Grant COMOTION (ERC-Küpper-614507), by the European Union's
Horizon 2020 research and innovation program under the Marie Skłodowska-Curie Grant Agreement 641789
``Molecular Electron Dynamics investigated by Intense Fields and Attosecond Pulses'' (MEDEA), and by
the Helmholtz Association through the Virtual Institute 419 ``Dynamic Pathways in Multidimensional
Landscapes'' and the ``Initiative and Networking Fund''. J.O.\ gratefully acknowledges a fellowship
by the Alexander von Humboldt Foundation.
\end{acknowledgments}


%
\end{document}